\documentclass[aps,prb,twocolumn,groupedaddress]{revtex4}
\usepackage{graphicx}
\usepackage{amssymb}

\begin{document}

\title{Dispersive force between dissimilar materials: geometrical effects}

\author{Cecilia Noguez}
\email[Corresponding author. Email:]{cecilia@fisica.unam.mx}
\affiliation{Instituto de F\'{\i}sica, Universidad Nacional Aut\'onoma
 de M\'exico, Apartado Postal 20-364, D.F. 01000,  M\'exico}

\author{C. E. Rom\'an-Vel\'azquez}
\affiliation{Instituto de F\'{\i}sica, Universidad Nacional Aut\'onoma
 de M\'exico, Apartado Postal 20-364, D.F. 01000,  M\'exico}

\date{\today}
\begin{abstract}
We calculate the Casimir force or dispersive van der Waals force between a spherical nanoparticle and a planar substrate, both with arbitrary dielectric properties. We show that the force between a sphere and a plane can be calculated through the interacting surface plasmons of the bodies. Using a Spectral Representation formalism, we show that the force of a sphere made of a material A and a plane made of a material B, differ from the case when the sphere is made of B, and the plane is made of A.  We found that the difference depends on the plasma frequency of the materials, the geometry, and the distance of separation between sphere and plane. The differences show the importance of the geometry, and make evident the necessity of realistic descriptions of the sphere-plane system beyond the Derjaguin Approximation or Proximity Theorem Approximation. 
\end{abstract}


\maketitle
\section{Introduction}
The origin of dispersive forces like Casimir \cite{casimir} and van der Waals (vW) forces \cite{pita,lifshitz},  between atoms and macroscopic bodies may be attributed to electromagnetic interactions between their charge distributions induced by quantum vacuum fluctuations, even when they are electrically neutral \cite{casimir,pita,lifshitz,milonni}. In a first approximation, the charge distribution of neutral particles may be represented by electric dipoles. This dipole approximation was employed by London to calculate the non-retarded van der Waals interaction potential $V_{\rm vW} (z)$ between two identical polarizable molecules by using perturbation theory in quantum mechanics\cite{london}.  Later, Casimir studied a simpler problem \cite{casimir}: the force between two parallel conducting plates separated by a distance $z$, due to the change of the zero-point energy of the classical electromagnetic modes. He found  an interaction energy per unit area, ${\cal V}_{\rm A}(z) = -(\pi^2 \hbar c/720)(1/z^3), $ where $c$ is the speed of light. In 1956, Lifshitz \cite{lifshitz} extended the theory of Casimir to dielectric semi-infinite slabs. A decade later, it was shown that the Lifshitz's formula could be rederived from the zero-point energy of the surface modes of the slabs \cite{kampen,gerlach,barash}. 

In recent years, accurate experiments~\cite{bressi,lamoraux,mohideen,chan,decca} have been performed to measure the Casimir force which is one of the macroscopic manifestations of the fluctuations of the quantum vacuum~\cite{casimir}. Most of these experiments~\cite{lamoraux,mohideen,chan,decca} have been done using a spherical surface and a plane, instead of two parallel planes, as originally proposed Casimir~\cite{casimir} himself. The interpretation of the Casimir force in these experiments is based on the Proximity Theorem which was developed by Derjaguin and collaborators~\cite{proximidad} to estimate the force between two curved surfaces of radii $R_1$ and $R_2$. The Proximity Theorem or Derjaguin Approximation (DA) assumes that the force on a small area of one curved surface is due to locally ``flat'' portions on the other curved surface, such that, the force per unit area is
\[ 
\mathcal{F}(z) = 2 \pi \left(\frac{R_1R_2}{R_1 + R_2}\right) {\mathcal V}(z), \label{ptf}
\]
where $\mathcal{V}(z)$ is the Casimir energy per unit area between parallel planes separated by a distance $z$. In the limit, when $R_1 =R$, and $R_2 \to \infty$, the problem reduces to the case of a sphere of radius $R$ and a flat plane, that yields to $\mathcal{F}(z) = 2 \pi R \mathcal{V}(z).$ The force obtained using the DA is a power law function of $z$, where at ``large'' distances is proportional to $z^{-3}$, while at short distances $ \mathcal{F}(z) \propto z^{-2}$. This theorem is supposed to hold when $z \ll R_1, R_2$, however, the validity of the DA has not been proved yet. 

In this direction, Johansonn and Apell \cite{appel} studied the vW force between a sphere and a semi-infinite slab. They calculated the electromagnetic stress tensor associated to the electric field correlation of the system from the Green's function of the problem expressed in bispherical coordinates. They concluded that for small separations  the behavior of the attractive force is consistent with the DA. However, one drawback of this formalism is that in bispherical coordinates the section surfaces become planar (a point) for small (large) values of $z$, so that, arbitrary values of ratio of $z/R$ cannot  be considered to study the sphere-plane system.

Casimir showed that the energy $\mathcal{U}(z)$, between two parallel perfect conductor planes can be found through the change of the zero-point energy of the classical electromagnetic field~\cite{casimir}, like
\begin{equation}
\mathcal{U}(z) = \frac{\hbar}{2} \sum_i [\omega_i(z) - \omega_i(z\to \infty)], \label{uint}
\end{equation}
where $\omega_i(z)$ are the proper modes that satisfy the boundary conditions of the electromagnetic field at the planes which are separated a distance $z$. For real materials, Lifshitz obtained a formula to calculate the Casimir force between parallel dielectric planes~\cite{lifshitz}. The force per unit area, according to the Lifshitz formula, is 
\[ 
f(z) = \frac{\hbar c}{2\pi^2} \int_0^\infty dQQ \int_{q\ge 0} dk \frac{k^3g}{q} {\rm Re} \frac{1}{\tilde k} \left[ \frac{1}{\xi^s -1} +  \frac{1}{\xi^p -1} \right],
\] 
with $\xi^\alpha = [r^\alpha_1 r^\alpha_2\exp(2i\tilde k z)]^{-1}$, where $r_j^\alpha$ is the reflection amplitude coefficient of the plane $j$ for the electromagnetic field with  $\alpha$-polarization. Here, $j=1$, or $2$, $\alpha = s$ or $p$, $q=\omega/c$, $\omega$ is the frequency, $c$ is the speed of light, $\vec{q} =(\vec{Q},k)$ is the vacuum wavevector with projections parallel to the surface $\vec{Q}$, and normal to the surface $k$, $\tilde{k} = k + i0^+$, and $g$ is the photon occupation number of the state $k$. 

The Lifshitz formula depends only on the reflection coefficients of the planes and the separation between them. This reflection coefficients can be calculated for arbitrary material using the surface impedance formalism for semi-infinite slabs~\cite{esquivel1:02,esquivel2:02,esquivel:03} and for finite slabs using, for example, the transfer matrix formalism~\cite{esquivel:01,esquivel3:02}. In addition, one observes that the Lifshitz formula is symmetric under the change of index $j= 1 \to 2$, it means that the force is indistinguishable under the change of planes which is natural given the symmetry of the system. 

The force between a sphere and a plane made of arbitrary dielectric materials is commonly calculated using the DA, where the Casimir energy $\mathcal{V}(z)$ is found using the Lifshitz formula. As a consequence, the force between a sphere and a plane is indistinguishable if the sphere is made of a material A and the plane is made of B, or if the sphere is made of B and the plane is made of A. Furthermore, the DA assumes that the interaction of a sphere and a plane can be calculated as the interaction of a plane, and a set of planes, and evaluates the force by adding contributions of various distances, as if the interaction between planes were independent. However, it is well known that the Casimir force is not an additive quantity \cite{milonni,reynaud}. 

In this paper, we show that the geometry of the system is important. We employ a method based on the determination of the proper frequencies of the system based on a spectral representation formalism \cite{ceci}. By calculating the zero-point energy of all the interacting surface plasmons of a sphere-plane configuration, we obtain that the force between dissimilar materials differs if the sphere is made of a material A and the plane is made of B, and vice versa. This provides an insight into the range of validity of the DA or Proximity Theorem. Also, we find a general formula to calculate the  Casimir force differences between arbitrary materials when the bodies are separated at least one radii of the sphere. In particular, we study the specific case of aluminum and silver. We restrict ourself to the nonretarded limit, this means, to the case of small particles at small distances. 

\section{zero-point energy of the interacting surface plasmons}

\subsection{Parallel planes}
In 1968, van Kampen and collaborators~\cite{kampen} showed that the Lifshitz formula in the nonretarded limit is obtained from the zero-point energy of the interacting surface plasmons of the planes. Later, Gerlach~\cite{gerlach} did an extension showing that also in the retarded limit the Lifshitz formula is obtained from the zero-point energy of the interacting surface plasmons. A comprehensive derivation of the Lifshitz formula using the surface modes is also found in Ref.~\onlinecite{milonni}. 

In the case of two parallel planes, in the absence of retardation, the proper electromagnetic modes satisfy the following expression~\cite{kampen} 
\begin{equation}
\left[ \frac{\epsilon (\omega )+1}{\epsilon (\omega )-1}\right] ^{2}\exp
[2kz]-1=0  \label{3}
\end{equation}
for any value of $0\leqslant k\leqslant \infty $. The roots $\omega_{i}(k) $ are identified as the oscillation frequencies of the surface plasmons. To find the proper modes, it is necessary to choose a model for the dielectric function of the planes. To illustrate the procedure for metallic planes, let us employ the plasma model for the dielectric function, as
\begin{equation}
\epsilon (\omega )=1- \frac{\omega_p^2}{\omega^2 }, \label{plasma}
\end{equation}
with $\omega_p$ the plasma frequency. Substituting Eq.~(\ref{plasma}) in Eq.~(\ref{3}), we find two proper modes that, at large distances ($kd\gg 1$), are given by
\begin{equation}
\omega_{\pm }\approx \frac{\omega_p}{\sqrt{2}}\left( 1\pm \frac{1}{2}\exp
[-kd]-\frac{1}{8}\exp [-2kd] \pm \cdots \right), 
\end{equation}
and in the limit $kd\rightarrow \infty $, we obtain $\omega_\pm = \omega_p/\sqrt{2}$ which are the frequencies of the surface plasmon of each half space. The zero-point energy per unit area will be given by
\begin{equation}
{\mathcal E}_{0}(z) =\frac{1}{2} \frac{\hbar \omega _{p}}{\sqrt{2}}2\pi \int_{0}^{\infty } \sum_i \omega_{i}(k) kdk , 
\end{equation}
Now, taking the difference of the zero-point energy when the planes are at a distance $z$, and when they are at infinite, we obtain the interaction energy of the system per unit area, like 
\begin{equation}
{\mathcal V}(z) = -\frac{\hbar \omega _{p}}{\sqrt{2}} \frac{\pi}{16z^2}.
\end{equation}
which depends on the energy of the surface plasmons of the half spaces and the distance between planes.
 
\subsection{Sphere and plane: dipolar approximation}
Now, let us consider the case of a sphere of radius $a$ and dielectric function $\epsilon_s(\omega)$, such that, the sphere is located at a minimum distance $z$ from the plane of dielectric function $\epsilon_p(\omega)$, as shown in Fig.~1. The quantum vacuum fluctuations will induce a charge distribution on the sphere which also induces a charge distribution in the plane. Then, the induced $lm$-th multipolar moment on the sphere is given by~\cite{claro}  
\begin{equation}
Q_{lm}(\omega)  =  \alpha_{lm} (\omega)\left[ V_{lm}^{\rm vac}(\omega) + V_{lm}^{\rm sub}(\omega) \right], \label{qlm}
\end{equation}
where $V_{lm}^{\rm vac}(\omega)$ is the field associated to the quantum vacuum fluctuations at the zero-point energy, $V_{lm}^{\rm sub}(\omega)$ is the induced field due to the presence of the plane, and $\alpha_{lm}(\omega)$ is the $lm$-th polarizability of the sphere. 
\begin{figure}[htb]
\centerline {
\includegraphics[width=3.15in]{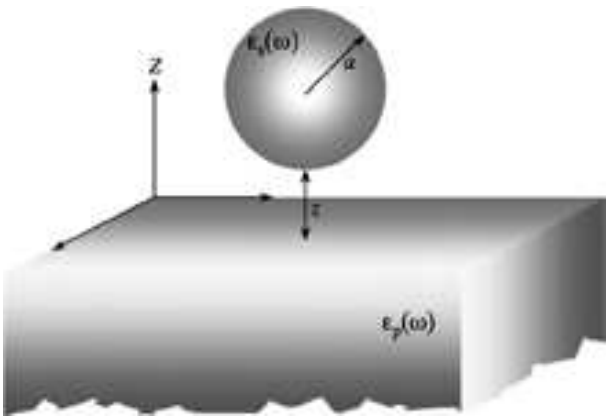}
}
\caption{Schematic model of the sphere-plane system.}\label{f1}
\end{figure}

To illustrate the procedure in detail, first we work in the dipolar approximation, i.e., when $l=1$. In section III, we will calculate the proper modes  taking into account all the high-multipolar charge distributions.
Using the method of images, the relation between the dipole moment on the the sphere ${\vec p}_{\rm s}(\omega)$, and the induced dipole moment on the plane ${\vec p}_{\rm p}(\omega)$, is
\begin{equation}
{\vec p}_{\rm p}(\omega) = \frac{1 - \epsilon_p(\omega)} { 1 + \epsilon_p (\omega)} {\mathbb M} \cdot {\vec p}_{\rm s}(\omega). \label{psub}
\end{equation}
Here, ${\mathbb M}= (-1, -1, 1)$ is a diagonal matrix whose elements depend on the choice of the coordinate system. From Eq.~(\ref{qlm}), one finds that the total induced dipole moment on the sphere is
\begin{equation}
{\vec p}_{\rm s}(\omega) =  \alpha(\omega) \left[ {\vec E}^{\rm vac} (\omega) + {\mathbb T} \cdot {\vec p}_{\rm p}(\omega) \right], \label{psph}
\end{equation}
where the interaction between dipoles is coupled by the dipole-dipole interaction tensor 
\[\mathbb T= (3 {\vec r} {\vec r} - r^2{\mathbb I})/r^5.\] 
Here, ${\mathbb I}$ is the identity matrix, and ${\vec r}$ is the vector between the centers of the sphere and the dipole-charge distribution on the plane. From  Fig.~1, we find that ${\vec r}=(0,0,2(z+a))$, such that, ${\mathbb M}\cdot {\mathbb T}= (-1/r^3, -1/r^3, -2/r^3)$ is a diagonal matrix. Substituting Eq.(\ref{psub}) in Eq.(\ref{psph}), one finds 
\begin{equation} 
\left[\frac{1}{\alpha(\omega)}\mathbb I + f_c (\omega) {\mathbb M}\cdot {\mathbb T} \right] \cdot {\vec p}_{\rm s} (\omega) = \mathbb G(\omega) \cdot {\vec p}_{\rm s} (\omega)= {\vec E}^{\rm vac} (\omega), \label{q2}
\end{equation}
where $\mathbb G (\omega)$ is a diagonal matrix and 
\[f_c (\omega)= [ 1 - \epsilon_{\rm p}(\omega)]/[  1 + \epsilon_{\rm p}(\omega)].\] 
Multiplying Eq.~(\ref{q2}) by $a^3$, one finds that each term of $\mathbb G(\omega)$ is dimensionless and has two parts: the left-hand side which is only associated to the material properties of the sphere through its polarizability $1/\tilde{\alpha}(\omega) = a^3/\alpha(\omega)$, and the right-hand side which is related to the geometrical properties of the system trough $a$, and $z$. The right-hand side also depends on the material properties of the plane trough the function $f_c(\omega)$.

The eigenfrequencies of the sphere-substrate system must satisfy Eq.~(\ref{q2}), and are independent of the exciting field, in this case ${\vec E}^{\rm vac}(\omega)$. These frequencies can be obtained when the determinant of the matrix in the left-side of Eq.~(\ref{q2}) is equal to zero, i.e., $\det \mathbb G (\omega) =0 $. In this particular case, the eigenfrequencies can be found from the following expression
\begin{equation}
\left[\frac{1}{\tilde{\alpha}(\omega)} +\frac{f_c(\omega)a^3 }{[2(z+a)]^{3}}\right]^2 
\left[ \frac{1}{\tilde{\alpha}(\omega)} +\frac{2f_c(\omega) a^3 }{[2(z+a)]^{3}}\right] = 0. \label{ceros}
\end{equation}
At this point, it is necessary to consider a model for the dielectric function of the sphere to find the proper modes. Again, we illustrate the procedure for  metallic sphere and  plane, using the plasma model of Eq.~(\ref{plasma}). Moreover, we use the fact that the polarizability of the sphere, within the dipolar approximation and in the quasi-static limit, is given by 
\[ \alpha(\omega) = a^{3} \frac{\epsilon_{\rm s}(\omega) -1} {\epsilon_{\rm s}(\omega) + 2}. 
\]

\begin{figure}[htb]
\centerline {
\includegraphics[width=3.1in]{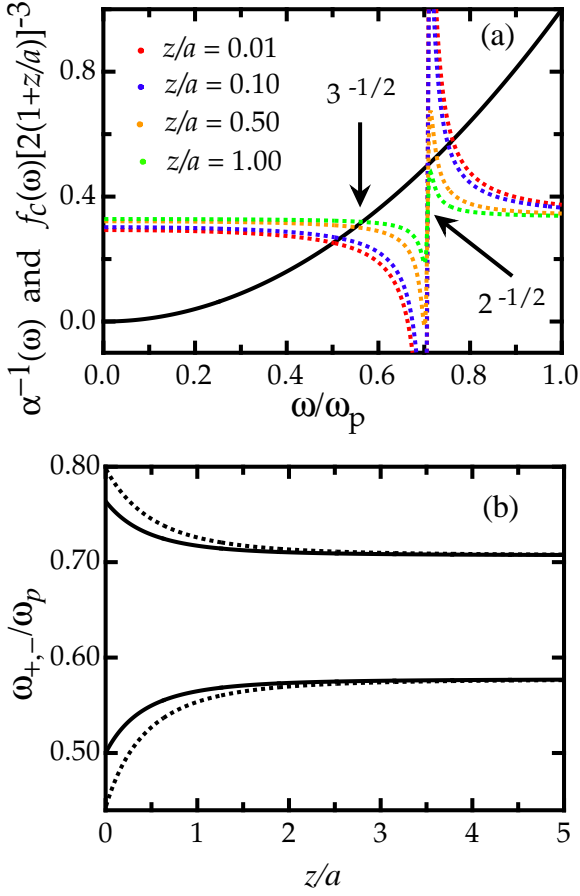}
}
\caption{(Color online) (a) Plot of $1/\tilde{\alpha}(\omega)$ in solid line and $f_c(\omega) [2(1+z/a)]^{-3}$ as a function of $\omega/\omega_p$ for different values of $z/a$. (b) Proper modes as a function of $z/a$.} \label{f2}
\end{figure}

In Fig.~\ref{f2}a, we plot, $1/\tilde{\alpha}(\omega)$ and $f_c(\omega) [2(1+z/a)]^{-3}$ as a function of $\omega/\omega_p$ for different values of $z/a$. The proper frequencies of the system are given when the black-solid line and the color-dotted lines intersect each other. We observe that there are two different proper modes for each term on Eq.~(\ref{ceros}), giving a total of six modes for each $z/a$. The proper modes at the left-hand side in Fig.~\ref{f2}a correspond to the surface plasmon of the sphere that we denote with $\omega_+$. These modes are red-shift as the sphere approaches the plane because of the interaction between them. When the separation $z/a$ increases, the modes $\omega_+$ go to $ \omega_p/\sqrt{3}$, which is the surface plasmon frequency of an isolated sphere. On the other hand, the proper modes at the right-hand side correspond to the surface plasmon of the plane and we denote them with $\omega_-$. These modes are blue-shift as the sphere approaches the plane, also because of the interaction between the sphere and the plane. In this case, when the separation $z/a$ increases, the mode $\omega_-$ goes to $\omega_p/\sqrt{2}$ which is the frequency of thesurface plasmon of the isolated plane. This behavior of the modes is clearly observed in Fig.~\ref{f2}b, where $\omega_{+}/\omega_p$ and $\omega_{-}/\omega_p$ are plotted as a function of $z/a$. This shows that the the zero-point energy of the sphere-plane system is directly associated to the interacting surface plasmons of the sphere and the plane.
 
\section{Exact Calculation of the zero-point energy of a sphere above a plane}

In this section we calculate  the proper electromagnetic modes of the sphere-plane system, within the nonretarded limit, by including all the high-multipolar interactions. These proper modes are calculated using a Spectral Representation formalism~\cite{fuchs,bergman,milton}, then, we calculate the zero-point energy using Eq.~(\ref{uint}). The Spectral Representation (SR) formalism uses the fact that the contributions of the dielectric properties of the sphere can be separated from the contributions of its geometrical properties. The latter allows to perform a systematic study of the sphere-plane system that may be very helpful for experimentalists. The details of the SR formalism can be found in Refs.~\onlinecite{ceci}~and~\onlinecite{ceci2}, here we only explain it briefly. 

We now calculate the proper modes of the system including all the high-multipolar charge distributions due to quantum vacuum fluctuations. Using the method of images, we find from Eq.~(\ref{qlm}) that the $lm$-th multipolar moment on the sphere is
\begin{eqnarray}
-\sum_{l'm'} \left[\frac{4 \pi \delta_{ll'} \delta_{mm'} }{(2l+1)\alpha_{l'm'}(\omega)} + f_c(\omega) A_{lm}^{l'm'}(z) \right] {{Q}}_{l'm'} (\omega)= && \nonumber \\
 V_{lm}^{\rm vac}(\omega),\quad &&  \label{q3}
\end{eqnarray}
where $A_{lm}^{l'm'}(z)$ is the matrix that couples the interaction between sphere and substrate~\cite{claro}. The expression for the interaction matrix $A_{lm}^{l'm'}(z)$ is given in the Appendix section, and it shows that $A_{lm}^{l'm'}(z)$ depends only on the geometrical properties of the system. If we have a homogeneous sphere, its polarizabilities are independent of the index $m$, and are given by~\cite{claro} 
\begin{equation}
\alpha_l (\omega) = \frac{l[\epsilon_{\rm s}(\omega) - 1]} {l [ \epsilon_{\rm s}(\omega) + 1] + 1} a^{2l+1} = \frac{n_{l0}}{n_{l0} - u(\omega)} a^{2l+1}, \label{alfa}
\end{equation}
where $n_{l0} = l/(2l+1)$, and $u(\omega) = [1 - \epsilon_s(\omega)]^{-1}$. Notice that in the above equation for the polarizability, the dielectric properties of the sphere are separated from its geometrical properties. By substituting the right-hand side of the above equation, we rewrite Eq.~(\ref{q3}) as:
\begin{eqnarray}
\sum_{l'm'} \Big\{ - u(\omega) \delta_{ll'}\delta_{mm'} + {H}^{l'm'}_{lm} (\omega,z) \Big\}  \frac{Q_{l'm'}(\omega)} { (l' a^{2l'+1})^{1/2}}  =&& \nonumber \label{multi} \\
 - \frac{(l a^{2l+1})^{1/2}} {4 \pi} V_{lm}^{\rm vac}(\omega), \quad &&
\end{eqnarray} 
where the $lm,l'm'$-th element of the matrix $\mathbb H(\omega,z)$ is
\begin{eqnarray}
H_{lm}^{l'm'} (\omega,z) = &&n_{l'0}  \delta_{ll'}\delta_{mm'} + \nonumber \\
						&& f_c(\omega) \frac{(a^{l+l'+1})} {4 \pi} (ll')^{1/2} A_{lm}^{l'm'}(z). \label{h}
\end{eqnarray}
The proper modes are given when the determinant of the quantity between parenthesis in Eq.~(\ref{multi}) is equal to zero, $\det{[- u(\omega) {\mathbb I} + \mathbb H (\omega,z)]}=0$, that is
\begin{equation}
G(\omega,z)=\prod_s [-u(\omega) + n_s(\omega,z)] =0, \label{ceros1}
\end{equation}
where $n_s(\omega,z)$ are the eigenvalues of $\mathbb H(\omega,z)$. 

The Spectral Representation formalism can be applied in a very simple way. First, we have to choose a substrate to calculate the contrast factor $f_c(\omega)$, with this, we construct the matrix $\mathbb H(\omega,z)$, and we diagonalize it to find its eigenvalues $n_s(\omega,z)$. We can repeat these steps for a set of different values of $z$. Notice that the eigenvalues of $\mathbb H(\omega,z)$ can be found  without to do any assumption on the dielectric function of the sphere. Once we have the eigenvalues as a function of $z$, we have to consider an explicit form of the dielectric function of the sphere, such that, we calculate the proper electromagnetic modes $\omega_s(z)$ trough the relation $u(\omega_s) = n_s(\omega,z)$. 

To illustrate the procedure, we use again the plasma model for the dielectric function of the sphere, and a plane with a constant dielectric function. In this case, the proper modes are given by
\begin{equation}
\omega_s(z) = \omega_p \sqrt{n_s(z)}, \label{modos2}
\end{equation}
with $s = (l,m)$, and where the eigenvalues $n_s(z)$, are independent of the frequency. The zero-point energy, according with Eq.~(\ref{uint}), is given by
\begin{equation}
{\cal E}(z) = \frac{\hbar \omega_p}{2} \sum_{l,m}\left[ \sqrt{n_{lm}(z)} - \sqrt{n_{l0}} \right],
\end{equation}
where $\hbar \omega_p\sqrt{n_{l0}} = \hbar \omega_p\sqrt{l/(2l+1)}$, are the energies associated  to the multipolar surface plasmons resonances of the isolated sphere. Here, $l = 1, 2, \dots, L$, where $L$ is the largest order in the multipolar expansion. For $l=1$ we have that $\hbar \omega_p\sqrt{n_{10}}= \hbar \omega_p/\sqrt3$, that is the surface plasmon of the sphere in the dipolar approximation. When $l \to \infty$, we have that $\hbar \omega_p\sqrt{n_{l0}} \to  \hbar \omega_p/\sqrt2$. 

Now, in the presence of the dielectric plane, the proper modes of the isolated metallic-sphere are modified according to Eq.~(\ref{h}). These proper modes are red-shift always as the sphere approaches the plane, and this shift depends on the separation $z/a$. We find that the interaction energy is proportional to $[1/2(1+z/a)]^{\beta}$, where $\beta = l+l'+1$ is an integer between $3$ and $2L + 1$.  As $z/a \to 0$, more and more multipolar interactions must be taken into account and $\beta \to 2L+1$. Finally, when the sphere is touching the plane,  one would expect that $L \to \infty$, so that also $\beta \to \infty$, and the energy at $z=0$ is proportional to $(1/2)^{\gamma}$, with $\gamma \to \infty$, it means the energy becomes also infinite.

It is possible to find the behavior of the energy as a function of $z/a$ for any plasma's sphere independently of its plasma frequency, like $\tilde{{\cal E}} \equiv {\cal E}/\hbar \omega_p $. Furthermore, the force can be also studied independently of the radius of the sphere, since ${\cal E}(z/a)$, one finds that 
\begin{equation}
{ F}=-\frac{\partial{\cal E}(z/a)}{\partial z} = -\frac{\partial{\cal E}(z/a)}{\partial(z/a)}  \frac{\partial(z/a)}{\partial z}, \label{force}
\end{equation}
and then, we can define also a dimensionless force like $a \tilde{ F} \equiv a {F} / \hbar \omega_p$. In summary, we have shown that given the dielectric properties of the substrate the behavior of the energy and force is quite general, showing the potentiality of the Spectral Representation formalism~\cite{ceci}. 

\begin{figure}[htp]
\centerline {
\includegraphics[width=3.1in]{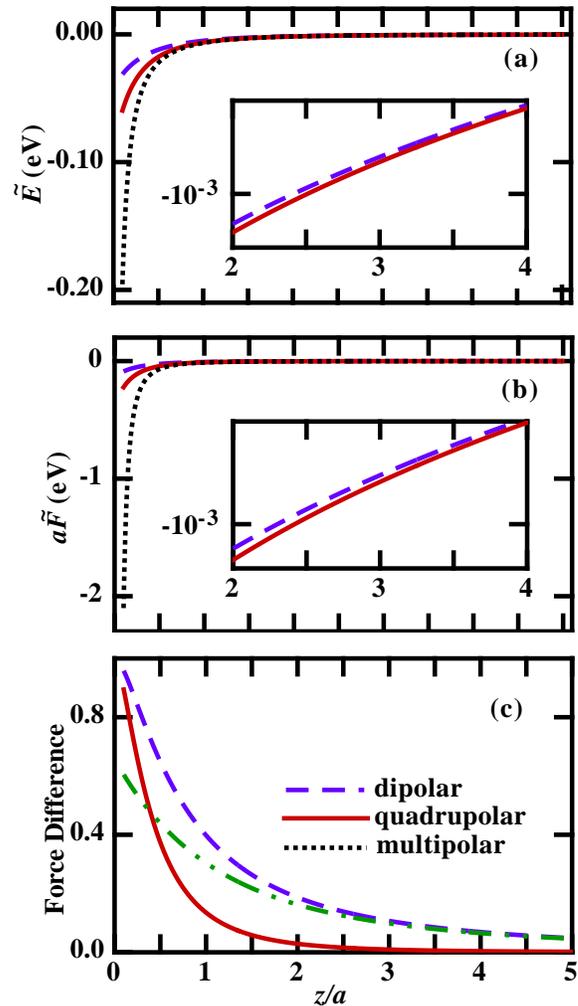}
}
\caption{(Color online) (a) Energy $\tilde{{\cal E}}$, and (b) force $a\tilde{F}$ as a function of $z/a$, with $L=80$ (dotted line), $L=2$ (solid line), and $L=1$ (dashed line). (c) Force difference, $\vert (\tilde{F}^{LH} - \tilde{F}^{LW})/ \tilde{F}^{LH} \vert$, using  multipolar moments $LH=80$ and $LW=2$ (solid line), $LH=80$ and $LW=1$ (dashed line), finally $LH=2$ and $LW=1$ (dot-dot-dashed line).}\label{f3}
\end{figure}

In Fig.~\ref{f3}, we show the dimensionless energy, $\tilde{{\cal E}}$, and force $a\tilde{F}$, for a sphere and a plane both described by the plasma model with the same plasma frequency. We show $\tilde{{\cal E}}$ and $a\tilde{F}$ as a function of $z/a$ when all multipolar interactions are taken into account, as well as only dipolar, and up to quadrupolar interactions are considered. The largest multipolar moment to achieve convergence of the energy at a minimum separation of $z/a=0.1$ is $L = 80$. We observe that $\tilde{{\cal E}}$ and $a\tilde{F}$ are power law functions of $(z/a)^{-\beta}$, where $\beta$ depends on the order of the multipolar interaction considered. Only within the dipolar approximation, $\tilde{{\cal E}}$ and $\tilde{F}a$ behaves like $(z/a)^{-3}$, and $(z/a)^{-4}$ for any $z$, respectively. Let us first analyze the system when only up to quadrupolar interactions are taken into account. In that case,  the energy as well as the force, show three different regions: (i) at large distances, $z > 5a$, only dipolar interactions are important, (ii) when $ 5a > z > 2a$, we found that dipolar-quadrupolar interactions become important and $\tilde{{\cal E}}$ and $\tilde{F}a$ behaves like $(z/a)^{-4}$, and $(z/a)^{-5}$; while (iii) at small distances $ z < 2a$ the quadrupolar-quadrupolar interaction dominates, and $\tilde{{\cal E}}$ and $\tilde{F}a$ behaves like $(z/a)^{-5}$, and $(z/a)^{-6}$, respectively

However, as the sphere approaches the substrate the interaction between high-multipolar moments becomes more and more important. The attractive force suddenly increases more than four orders of magnitude, as compare with the dipolar and quadrupolar approximations. In Fig.~\ref{f3}c we show the difference of the dimensionless force, $\vert (\tilde{F}^{LH} - \tilde{F}^{LW})/ \tilde{F}^{LH} \vert$, as a function of the separation, with $LH$ and $LW$ the highest and lowest-multipolar moments taken into account. We observe that at distances $ z > 2a$ the force can be obtained if up to quadrupolar interactions are considered. We also found that for $z > 7a$ the interaction between the sphere and the substrate can be modeled using the dipolar approximation only, like in the Casimir and Polder model~\cite{casypol}. However, at separations smaller than $z < 2a$, the quadrupolar approximation also fails, and it is necessary to include high-multipolar contributions. For example, at $z=a$ the quadrupolar approximation gives an error of about $10\%$, while the dipolar approximation gives an error of about $40\%$. At smaller distances like $z=a/2$, the quadrupolar approximation gives an error of about $40\%$, while in the dipolar approximation the error is $70\%$. At $z=0.1a$ both approximations give an error larger than $90\%$. To directly compare the dipolar and quadrupolar approximations, we also plot in Fig.~\ref{f3}c the force difference when $LH=2$ and $LW=1$. In this curve, we clearly observe that even at $z=5a$ there are differences between the dipolar and quadrupolar approximations of about $5\%$. 

The large increment of the force at small separations due to high-multipolar effects could explain the physical origin of, for example, the large deviations observed in the deflection of atomic beams by metallic surfaces~\cite{shih}, as well as some instabilities detected in micro and nano devices. Furthermore, to compare experimental data of some experiments~\cite{chan,decca:2} with the DA, it has been needed to make a significant modification to the Casimir force. In Ref.~\onlinecite{chan} the authors found a larger force than the one calculated using DA, while in Ref.~\onlinecite{decca:2} the authors measured a smaller one. Both groups argue that the discrepancies between theory and experiment are due to the same effect: the roughness of the surface. However, the discussion around this point has not sense until the DA be proven. Indeed, this ambiguity  on the interpretation of the measurements indicates that the force between a spherical particle and a planar substrate involves more complicated interactions than the simple dipole model~\cite{casypol} or the DA~\cite{proximidad}. For example, such deviations might be attributed also to high-multipolar effects, however, new experiments have to be perform to prove the later.

In this paper, we propose a way to elucidate experimentally two important issues to the Casimir force: (i) the relevance of the geometry via the surface plasmon interactions between bodies, and (ii) the relevance of the high-multipolar interactions in the sphere-plane model. In the next section, we study the case of the casimir force between a sphere and a plane of dissimilar materials that can help us to clarify these important issues. 

\section{Casimir force between dissimilar materials}

Using the formalism presented in the previous section, we calculate the Casimir force for a sphere and a plane made of dissimilar materials. To illustrate our results, we choose silver and aluminum whose dielectric function is described by the Drude model 
\begin{equation}
\epsilon (\omega )=1- \frac{\omega_p^2}{\omega(\omega + i/\tau)}, \label{drude}
\end{equation}
with plasma frequencies $\hbar \omega_p = 9.6$~eV, and $15.8$~eV, and relaxation times $(\tau \omega_p)^{-1} = 0.0019$, and $0.04$, respectively. In this case, we employ the Argument Principle (AP) to calculate the force. The AP was first used by van Kampen~\cite{kampen} and Gerlach~\cite{gerlach} to calculate the force between planes. The reader can consult Ref.~\onlinecite{milonni} for a comprehensive explanation of the use of the AP within the context of Casimir forces. 

Then, employing the Argument Principle method we find that the sum over proper modes is given as
\begin{equation}
\label{ener}
{\mathcal E}(z) =\sum_s \frac{\hbar \omega_s(z)}{2} = \frac{\hbar}{4 \pi i } \oint_C \omega \frac{\partial G(\omega,z)}{\partial\omega}\frac{1}{G(\omega,z)} d\omega \, ,
\end{equation}
where $G(\omega,z)$ is defined in Eq.~(\ref{ceros1}), such that 
\begin{equation}
{\mathcal E}(z)  = \frac{\hbar}{4\pi i } \sum_s \oint_C \omega \frac{\partial}{\partial \omega}\log{[-u(\omega) + n_s(\omega,z)]}  d\omega .
\end{equation}
The closed contour $C$ is defined as the part along the imaginary axis plus the part along the semicircle on the positive real axis. In the limit for a infinite semicircle radius, the integral along the semicircle does not depend on the separation $z$, therefore, it does not contribute to the force. Then, it follows that the interaction energy of Eq.~(\ref{uint}) is given by
\begin{equation}
{\mathcal U}(z) =\frac{-\hbar}{4\pi i } \sum_s \int_{-i\infty}^{i\infty} d\omega \log{[-u(\omega) + n_s(\omega,z)]}, \label{ap}
\end{equation}
where in the last step we have performed a partial integration. The force is found as in Eq.~(\ref{force}), where we first perform the derivate with respect to $z$ of Eq.~(\ref{ap}) and then the integral with respect to $\omega$. 

\begin{figure}[hbt]
\centerline {
\includegraphics[width=3.1in]{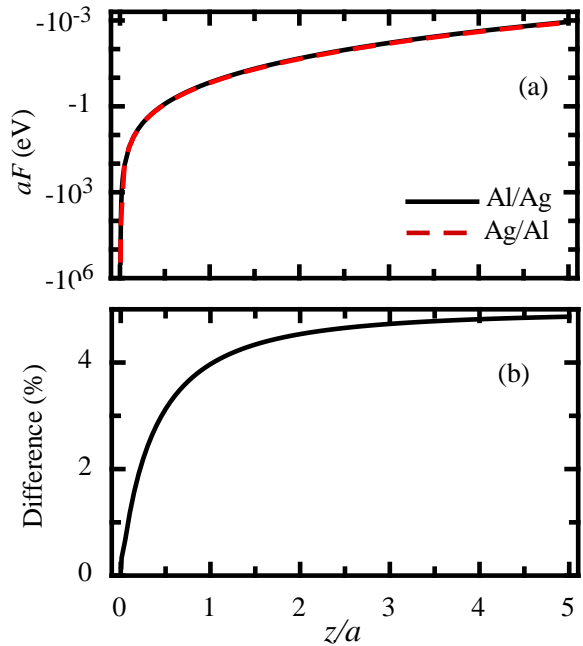}
}
\caption{(Color online) (a) Force as a function of $z/a$, for a sphere made of Al over an Ag plane (solid line), and for a sphere made of Ag over an Al plane (dashed line). (b) Difference of the force between a sphere made of Al over an Ag plane, and the force for a sphere made of Ag over an Al plane.} \label{f4}
\end{figure}

In Fig.~\ref{f4}, we show the calculated force for an aluminum sphere of arbitrary radius above a silver plane, $aF_{\rm Al/Ag}$,  in solid line, and the same for a gold sphere above an aluminum plane, $aF_{\rm Ag/Al}$, in dashed line. We find that the difference is small and cannot be directly observed. Then, in Fig.~\ref{f4}b we show the difference of the force normalized with the average as a function of $z/a$, given by
\begin{equation} 
\Delta \mathcal F = 2\left \vert \frac{F_{\rm A/B} - F_{\rm B/A}}{ F_{\rm A/B} +  F_{\rm B/A}} \right\vert,
\end{equation}
with A=Al, and B=Ag. Here, we observe that the difference of force varies as a function of $z/a$, such that at small distances, ($z < 0.5 a$) the difference under the interchanged of materials between the sphere and the plane, is less than $3\%$. Then, the difference increases with the minimum separation, and at $z > a$, it has almost reached their maximum. For these particular materials, the maximum is about $4.86\%$, independently of the radius of the sphere. We find that this value of $\Delta \mathcal F $ is a consequence of the geometry of the system, as we explain below.

In the previous section, we show that for a minimum separation larger than $2a$, the force can be described with quadrupolar interactions between the sphere and plane. This means that for $z> 2a$ the force varies with the same power-law function. This conclusion can be easily deduce from Fig.~\ref{f3}c. We also show in section~II, that the Casimir force is given by the interaction of the surface plasmons of the sphere and plane. In Fig.~\ref{f2}b, we observe that the proper modes of the system for $z > a$ are approximately given by $\hbar \omega_p/ \sqrt{2}$ and $\hbar \omega_p/ \sqrt{3}$, when $L=1$. These yield to conclude that for $z > a$, the force difference is independent of $z$, such that
\begin{eqnarray} 
\Delta {\mathcal F} &\simeq & 2 \left \vert \frac{\left(\frac{\hbar \omega_p^{\rm A}}{\sqrt{3}} +  \frac{\hbar \omega_p^{\rm B}}{\sqrt{2}} \right) - \left(\frac{\hbar \omega_p^{\rm B}}{\sqrt{3}} +  \frac{\hbar \omega_p^{\rm A}}{\sqrt{2}} \right)}{\left(\frac{\hbar \omega_p^{\rm A}}{\sqrt{3}} +  \frac{\hbar \omega_p^{\rm B}}{\sqrt{2}} \right) + \left(\frac{\hbar \omega_p^{\rm B}}{\sqrt{3}} +  \frac{\hbar \omega_p^{\rm A}}{\sqrt{2}} \right)} \right \vert, \nonumber \\
& = & 0.202 \left \vert \frac{\hbar \omega_p^{\rm A} -\hbar \omega_p^{\rm B}}{\hbar \omega_p^{\rm A} + \hbar \omega_p^{\rm B}} \right\vert . \label{diff}
\end{eqnarray}
For the case of A=Al, and B=Ag, we find from Eq.~(\ref{diff}) that $\Delta {\mathcal F} \simeq  4.9 \% $, which is the calculated value shown in Fig.~\ref{f4}b when $z > 2a$. When $z <a$ the difference decreases as $z/a$ also does, because more and more multipolar interactions become important as the sphere approaches the plane. In the case of Al and Ag, the force difference at $z/a = 0.001$ is less than $0.01\%$, while at $z/a = 0.1$ the difference is about $1.2\%$. In Eq.~(\ref{diff}) we observe that  as the plasma frequency of the materials becomes more dissimilar the difference of the force becomes larger. For example, for potassium ($\hbar \omega_p = 3.8$~eV) and aluminum we would find that $\Delta {\mathcal F} \simeq 12.4\%$ when $z> 2a$, while for gold ($\hbar \omega_p = 8.55$~eV) and copper ($\hbar \omega_p = 10.8$~eV), $\Delta {\mathcal F} \simeq 2.4\%$ when $z > 2a$. In the other hand, the differences become very small when $z < 0.1$ independently of the materials because of the high-multipolar surface plasmons interactions.

\section{Summary}
In this work, we show that the Casimir force between a sphere and a plane can be calculated through the energy obtained from their interacting surface plasmons. Our result is in agreement with the work of van Kampen and collaborators~\cite{kampen}, and Gerlach~\cite{gerlach} that showed that the Casimir force between parallel planes can be obtained from the zero-point energy of the interacting surface plasmons of the planes. We show that the Casimir force of a sphere made of a given material A and a plane made of a given material B, is different from the case when the sphere is made of B, and the plane is made of A. We obtain a formula to estimate the force difference for Drude-like materials, when they are separated at least two radii of the sphere. We find that the difference depends on (i) the plasma frequency of the materials, and (ii) the distance of separation between sphere and plane. We also found that as the sphere approaches the plane, the force difference becomes smaller. We conclude that the geometry of the system is important at distances larger than the radii of the sphere, where the force is dominated by the surface plasmons of the bodies. The energy of these surface plasmons is equal to the plasma frequency of the materials times a factor given by the geometry.

\section*{Acknowledgments}
We would like to thank to Carlos Villarreal, Ra\'ul Esquivel-Sirvent and Rub\'en G. Barrera for helpful discussions. This work has been partly financed by CONACyT grant No.~36651-E and by DGAPA-UNAM grant No.~IN104201.
 
\begin{widetext}
\appendix*
\section*{Appendix}
The tensor that couples the interaction between the sphere and the plane is~\cite{claro}
\begin{equation}
A_{lm}^{l^{^{\prime }}m^{^{\prime }}} = \frac{ Y^{m-m^{^{\prime }}}_{l+l^{^{\prime }}} (\theta, \varphi)} {r^{l+l^{^{\prime }}+1} } 
\left[ \frac{(4\pi)^3 (l+l^{^{\prime }}+m-m^{^{\prime }})!(l+l^{^{\prime }}-m+m^{^{\prime }})!} {(2l+1) (2l^{^{\prime}}+1)(2l+2l^{^{\prime }}+1)(l+m)!(l-m)!(l^{^{\prime }}+m^{^{\prime}})! (l^{^{\prime }}-m^{^{\prime }})!} \right]^{1/2}
\end{equation} 
where $Y_{lm}(\theta, \varphi)$ are the spherical harmonics. In spherical coordinates ${\vec r}= (2(z+a), 0,0)$, then, the spherical harmonics reduce to~\cite{arfken}
\begin{equation}
Y^{m-m^{^{\prime }}}_{l+l^{^{\prime }}} (0, \varphi) = \left[ \frac{2(l+l^{^{\prime }}) +1}{4 \pi} \right]^{1/2} \delta_{m-m^{^{\prime }}, 0}.
\end{equation}
This means that $m=m^{'}$ and multipolar moments with different azimuthal charge distribution are not able to couple or interact among them. This yields to
\begin{equation}
A_{lm}^{l^{^{\prime }}m^{^{\prime }}} = \frac{4 \pi}{[2(z+a)]^{l+l'+1}} \left[ \frac{ ((l+l')!)^2}{(2l+1) (2l^{^{\prime}}+1) (l+m)! (l-m)! (l^{^{\prime }}+m)! (l^{^{\prime }}-m)!} \right]^{1/2}, 
\end{equation}
which is a symmetric matrix. 
\end{widetext}

\bibliography{prb-dis}





\end{document}